\documentstyle[preprint,prl,aps]{revtex}	

\begin{document}

\draft  					

\title{Quantum melting on a lattice and a delocalization transition}  
\author{E. V. Tsiper,$^1$ F. G. Pikus,$^2$ A. L. Efros$^1$}
\address{$^1$ Department of Physics, University of Utah, Salt Lake
City,
UT 84112}
\address{$^2$ Department of Physics, University of California at Santa
Barbara, Santa Barbara, CA 93106}
\date{\today}
\maketitle

\widetext

\begin{abstract} We consider 2$d$ gas of spinless fermions
with the Coulomb interaction on a lattice at $T=0$ and at different
values of the hopping amplitude $J$.  At small $J$ electrons form a
periodic structure.  At filling factor $\nu=1/6$ this structure melts
at $J$ as low as 0.02--0.03 in units of the nearest-neighbor Coulomb
energy.  We argue that this transition is connected to the
dielectric-metal or dielectric-superconductor transitions.  To
demonstrate this point we perform computer modeling of the systems
$6\times 6$ with 6 and 7 electrons and $6\times 12$ with 12 electrons.
By sweeping $J$ we compute simultaneously persistent current and
structural characteristics of the electron distribution.
\end{abstract}


\narrowtext

The great majority of the efforts made recently to study correlated
particles on a lattice were restricted to the Hubbard model or $t-J$
model (See review \onlinecite{dag}).  We concentrate here on a direct
interaction between particles which introduces some important new
physical features.  We consider spinless fermions at $T=0$ on a
2-dimensional square lattice with the following Hamiltonian
\begin{equation}
H=J\sum_{{\bf r},{\bf s}}a_{{\bf r}+{\bf s}}^\dagger a_{\bf r}
  \exp(i\bbox{\phi s})
+ \frac{1}{2}\sum_{{\bf r}\neq{\bf r}^{'}}
  \frac{n_{\bf r}n_{{\bf r}{'}}}
  {|{\bf r}-{\bf r}^{'}|}.
\end{equation}

Here $n_{\bf r}=a_{\bf r}^\dagger a_{\bf r}$, the summation is
performed over the lattice sites {\bf r}, {\bf r}$^{'}$ and over the
vectors of translations to the nearest-neighbor sites, {\bf s}.  The
lattice constant and the Coulomb energy between nearest neighbors are
chosen to be the length and energy units.  We use periodic conditions
at the boundaries of rectangle $L_x\times L_y$.  The dimensionless
vector potential $\bbox{\phi}=(\phi_x,\phi_y)$ in the Hamiltonian is
equivalent to the twist of the boundary conditions by the flux
$\Phi_i=L_i\phi_i$, $i=x,y$.  The total spectrum is periodic in
$\Phi_x$ and $\Phi_y$ with the period $2\pi$.

In a commensurate phase, at small enough values of $J$, fermions form
a Wigner crystal (WC) with the long range order.  At $T=0$ and some
critical value $J_c$ the crystal melts.  Not much is known about the
quantum melting on a lattice.  The only exactly solvable
model\cite{1d} is a 1-d model with the nearest-neighbor interaction
and with the filling factor $\nu=1/2$.

In 2-d case a computer modeling made by Pikus and Efros\cite{pik} at
$\nu=1/3$ and $1/6$ in a square $6\times 6$ shows that $J_c$ is as low
as 0.03--0.02.  They argue that the lifting of degeneracy of the
ground state is the best diagnostic of the transition.  At $J=0$ this
is an exact degeneracy corresponding to the different translational
positions of a crystal on a background lattice.  In a finite array it
is lifted at any small $J$ but the splitting is proportional to
$\exp(-N|\ln J|)$, where $N$ is the total number of the particles.
Above $J_c$ one should expect a macroscopic splitting.
 
Since the Hamiltonian (1) is translationally invariant, the states of
the system can be classified in terms of the total quasimomentum {\bf
P}.  The degeneracy of the ground state at $J<J_c$ means that the
energy of the macroscopic system is independent of {\bf P}.

We show here that the structural transition is accompanied not only by
the {\bf P}-splitting of the ground state. It includes also a {\em
delocalization}, defined as an appearance of the {\em persistent
current}, i.\ e.\ dependence of the energy on the flux $\Phi$.  We
have found that at small $J$ the system is dielectric and at $J\approx
J_c$ it comes to a new state, which can be either metallic or
superconducting.

The WC in a continuous media without any disorder has the same flux
response as free electrons\cite{krive}.  The dielectric behavior of
the lattice WC at small $J$ can be understood in terms of the umklapp
processes.  Yet another explanation is also possible which resembles
the ideas of Ref.\ \onlinecite{kagan}.  Consider Hamiltonian Eq.\ (1)
for two particles only.  If the distance between the particles is such
that their interaction energy is larger than the width of the band,
which is $8J$, the particles form a bound state.  They cannot be
separated because of the energy conservation law.  The center-of-mass
motion for these two particles is allowed, but corresponding band has
a width proportional to $J^2$ at small $J$.  In the same way $n$
interacting particles will also form a bound state with the bandwidth
of the center-of-mass motion proportional to $J^n$.  We may then
conclude that a system of interacting particles blocks itself if an
interaction energy at {\em average distance} is much larger than the
width of the band.  Note that this second interpretation is applicable
not only to commensurate phases\cite{els}.  The transition we are
considering is of a correlation nature.  But it differs from the
Mott-Hubbard transition because it occurs due to the interaction of
remote electrons and because spin does not play any important role
here.

As a basis for computations we use many-electron wave functions at
$J=0$.  Their total number is $C_M^N$, where $M=L_x\times L_y$ is the
area of a system.  They can be visualized in a form of pictures, which
we call {\em icons}.  Some icons with the lowest energies are shown in
Fig.\ 1.  The lowest energy has the icon with a fragment of the
crystal.  Icons with larger energies may also represent periodic
structures with a period less than the size of the system.  The
Coulomb energy has been calculated as a Madelung sum, assuming that
the icons are repeated periodically through the infinite plane with a
compensating homogeneous background.  Each icon represents a Slater
determinant $\Psi_\alpha$.

For each icon $\alpha$ there are $m_\alpha$ different icons that can
be obtained from it by various translations.  These icons are combined
to get the wave function with total quasimomentum {\bf P}:
\begin{equation}
  \Psi_{\alpha{\bf P}}=\frac{1}{\sqrt{m_\alpha}}
  \sum_{\bf r}\exp(i{\bf P}{\bf r})T_{\bf r}\Psi_\alpha.
\end{equation}
The summation is performed over $m_\alpha$ translations $T_{\bf r}$.
Important point is that the icons with periodic structures generate
smaller number of different functions $\Psi_{\alpha{\bf P}}$.  The
number of allowed {\bf P} generated by each icon is equal to
$m_\alpha$.  In particular, for each icon of a WC with one electron
per primitive cell, one has $m_0=1/\nu$.

The allowed values of {\bf P} generated by a WC icon are determined
from the conditions $(-1)^{Q_j}\exp(i{\bf P}{\bf l}_j)=1$.  Here ${\bf
l}_j$ are the primitive vectors of the WC, and $Q_j$ are the numbers
of fermionic transmutations necessary for translations on these
vectors.  These conditions can be easily understood.  If translation
on a vector ${\bf l}_j$ is applied to Eq.\ (2), the right-hand side
acquires a factor $(-1)^{Q_j}$, while for a function with given {\bf
P} this factor must be equal to $\exp(i{\bf P}{\bf l}_j)$.  If $Q_j$
are even for both ${\bf l}_j$, the allowed {\bf P} form the reciprocal
lattice of the WC.  However, in the case when one or both of $Q_j$ are
odd, the lattice is shifted by $\pi$ in the corresponding directions.
In such case ${\bf P}=0$ is forbidden.  The set of $m_\alpha$
nontrivial values of {\bf P} is restricted to the first Brillouin zone
of the background lattice.  One WC is represented by a number of icons
obtained from each other by the point-group transformations of the
background lattice.  The total number of allowed values of {\bf P} for
the WC is the property of the WC and does not depend on the size and
the shape of the system.  Contrary, an icon representing a point
defect in a WC generates all vectors {\bf P}; their total number is
$M$.

The following results can be obtained directly from the perturbation
theory with respect to $J$: 1.  The ground state and the lowest
excited states have a common large down shift which is proportional to
$J^2$ and to the total number of particles $N$.  2. The ground state
splitting appears in the $N$-th order and it is proportional to $J^N$.
3. The flux dependence of the ground state for the flux in
$x$-direction appears in the $L_x$-th order and it is proportional to
$J^{L_x}$.

The transition we are studying occurs at such small values of $J$ that
classification of states in terms of icons is still useful.  At small
$J$ there is a gap in the spectrum since a finite energy is required
to create a point defect in the WC.  The states originating from the
WC icon do not belong to the continuous spectrum since their number
remains finite in a macroscopic system.  The states originating from
an icon with point defect in WC form a band.

One can imagine two different scenario of the transition.  The
simplest one is the first-order phase transition.  It occurs if the
branch originated from the point-defect icon crosses the ground state
at $J=J_c$.  This may happen because the energy of the bottom of the
point-defect band is going down with increasing $J$ and can overcome
the Coulomb energy of the point defect existing at $J=0$.  The
crossing is possible if the defect branch has {\bf P} different from
all vectors of the WC.  Since point defect may have all {\bf P}, the
excitation spectrum of the large system will become continuous at
$J>J_c$.  Then the new state should be a normal metal.

In the second scenario the ground state eigenvector originated from
the WC icon has an avoided crossing with defect states of the same
{\bf P}.  The ground state obtains a large admixture of the defect
states, looses the structural long range order and becomes delocalized
in terms of the persistent current.  However, the gap between the
ground state and an excited state remains finite.  This would be the
second-order transition and the resulting state might be
superconducting.

We have performed numerical diagonalization of the Hamiltonian for the
system $6\times 6$ with 6 and 7 electrons, and for the system $6\times
12$ with 12.  The exact-diagonalization results for the system with 6
electrons are shown in Figs.\ 2 and 3.  For 12 electrons the complete
basis consists of $1.5\times 10^{13}$ functions and the exact
diagonalization is impossible.  We performed diagonalization with a
truncated basis as proposed in Ref.\ \onlinecite{pik}.  Two sets of
$2.9\times 10^7$ and of $5.8\times 10^7$ different icons have been
created by the classical Monte-Carlo with the temperatures of
generation\cite{pik} $T_g=0.020$ and $0.025$ correspondingly.  These
sets make approximately $1.9\times 10^{-5}$ and $3.8\times 10^{-5}$ of
the complete basis.  For each {\bf P} the size of the matrix was
$4\times 10^5$ and $8\times 10^5$ respectively.  The flux
sensitivities $S$ calculated by diagonalization of these matrices are
shown in Fig.\ 3 by open and solid symbols.  Except a small region at
$J\approx0.03$ there is no visible difference between the results.

For 7 electrons we are able to check how the computations with
truncated bases converge to an exact result\cite{els}.  We obtained
the following empiric rule: if one takes 2 truncated basic sets, one
twice as large as the other, the difference in the results gives an
upper boundary for the error for the largest set.  This rule is
applicable to the computation of the total energy as well as for {\em
difference} in energy caused, say, by flux.  It works when the
relative error is small.

We think the picture we observe in a system with 6 electrons resembles
the second scenario proposed above since the defect branch never
crosses the ground state.  Fig.\ 2 shows a few lowest energies at
different momenta {\bf P} as a function of $J$ at zero flux.  At $J=0$
we get the Coulomb energies of icons. The energy of the lowest icon is
taken as zero. The second order perturbation theory shift is
calculated using the energies of icons to be $-177J^2$, and it is
subtracted from the total energy.  The state with ${\bf P}=(0,\pi)$ is
the lowest at $J<0.02$.  At larger $J$ the lowest state has ${\bf
P}=(\pi/3,\pi/3)$ and it also originates from the WC icon.

To study delocalization we compute a flux sensitivity $S(\Phi)=
E(\Phi)-E(0)$, the difference between the ground state energies with
and without flux $\Phi$.  To give more information we plot in Fig.\ 3
the energy difference for two above-mentioned values of {\bf P} for
all $J$.  Each of these curves represents $S(\Phi)$ in the region of
$J$ where the branch with corresponding {\bf P} is the ground state.
Fig.\ 3 shows data for $\bbox{\phi}||{\bf P}$ and $\bbox{\phi}\bot
{\bf P}$, $\Phi_{x,y}$ either 0 or $\pi$.  One can see a sharp
increase of $S$ with $J$ for both values of {\bf P} and both
directions of $\bbox{\phi}$. In agreement with the perturbation theory
it is proportional to $J^6$ at $J\leq0.02$.

The interpretation of the results for $6\times 12$ system is more
ambiguous.  As one can see from Fig.\ 1 (e)--(h), the low-energy part
of the spectrum is created by three icons with close energies.  The
lowest one represents the WC.  Two others can be classified neither as
a crystal nor as a point defect.  They represent rather interdomain
boundaries in the WC.  At $J\approx 0.004$ the branch with ${\bf
P}=(0,\pi/3)$ originated from the second icon becomes the ground
state.  Such {\bf P} is not allowed for the WC.  Therefore, the
situation resembles the first scenario leading to the phase transition
of the first order.  We think, however, that the small Coulomb gap
between these icons, and, as a result, small $J_c$ should be
considered as an artifact of a small system.  Since in a macroscopic
system the energy of an interdomain boundary is proportional to the
size of the system, such icons do not contribute to the lowest part of
the energy spectrum.  In smaller $6\times 6$ system such a boundary
does not even appear because of the periodic conditions.

That is why we study the states originated from the WC icon ignoring
that none of them is the ground state.  The lowest state has ${\bf
P}=(0,\pi)$ as in the case of $6\times 6$ system.  The functions
corresponding to the 3 lowest icons have a very small admixture of
each other and can be studied separately.  The flux sensitivity $S$ of
the lowest ${\bf P}=(0,\pi)$ branch is shown in Fig.\ 3 by open and
solid circles.  The sharp step-like increase in $S$ is the result of a
crossing of two such branches.  By open and solid squares we also show
$S$ for the branch with the largest admixture of the WC icon having in
mind that due to the above arguments this branch might become the
ground state in a macroscopic system.  Note again, that the same
branch with ${\bf P}=(0,\pi)$ is the ground state of the $6\times 6$
system.

For the same WC branch in the system $6\times 12$ we have calculated
the correlation function $G({\bf r}-{\bf r}^{\prime})=\left<n_{\bf
r}n_{{\bf r}^{\prime}}\right>$.  At very small $J$ the function
$G(\bbox{\rho})$ is close to 1 at all vectors $\bbox{\rho}$ of the WC
and close to zero otherwise.  To demonstrate melting we plot in Fig.\
3 the difference in $G$ for two equivalent WC vectors (3,0) and (0,6)
as a function of $J$.  The difference is taken to cancel $J^2$-term,
which is the same for all equivalent points and does not show the loss
of the long range order.  One can see that the structural
transformation appears in the same range of $J$ as the delocalization.

The dependence $S(\Phi)$ shown in Fig.\ 4 for both 6 and 12 electrons
at $J=0.04$.  In the first case this is the difference of two ground
state energies, in the second case this is the change of energy in the
WC branch.  In both cases we have found a very strong admixture of the
second harmonic, which dominates for the system with 12 electrons.
This means an admixture of defects with double charge and it gives
another argument that the state after transition might be
superconducting.  The admixture is strong in the region $0.02<J<0.04$
and it becomes less both with decreasing and increasing $J$.

 The value of $S$ after the transition increases linearly with $J$ and
is of the order of $J$. This result agrees with an estimate from the
London equation which gives $S\sim \nu m^{-1}\sim \nu J$, where
$m$ is an electron mass.  The large magnitude of $S$ is a result of a
gap in the spectrum. The most interesting question is whether this gap
survives in a macroscopic sample. We think this may happen because the
gap originates from the excitation energy of a defect, and all
relevant characteristics such as the width of the defect band and the
matrix element of avoiding crossing with the ground state should not
have a strong size dependence.

Finally, we have shown that strong electron correlations create a
dielectric state in 2d gas with a narrow band, and we have studied the
collapse of this state with increasing band width.  Our data for
$\nu=1/6$ are more in favor of the dielectric-superconductor than the
dielectric-metal transition.

We are grateful for useful discussions to L. P. Gor'kov, D. Mattis, E.
I. Rashba, B. Z. Spivak, and Bill Sutherland.  The work is supported
by QUEST of UCSB, subagreement KK3017 and by SDSC.  F. G. P.
acknowledges support by grant NSF DMR 9308011.


\newpage
\centerline{FIGURES}

FIG.\ 1.  Four icons with the lowest Coulomb energies for each of the
two systems at $\nu=1/6$.  The total Coulomb energies $E$ are shown
above each icon.  The lowest-energy icons (a,e) represent fragments of
the same Wigner crystal.

FIG.\ 2.  The low-energy part of the spectrum of the system $6\times
6$ with 6 electrons as a function of $J$.  The energy is measured from
the Coulomb energy for the WC icon, $E_0=-4.75547$.  Branches with
different quasimomentum {\bf P} are shown with different lines.  The
numbers in brackets show the components of the vector {\bf P} in units
$\pi/3$.

FIG.\ 3.  Flux sensitivity $|S(\pi)|$ for different branches of the
spectrum.  The curves for the system $6\times 6$ with 6 electrons are
the result of exact diagonalization.  The open and solid symbols show
the results of diagonalizations of two truncated basic sets for the
system $6\times 12$ with 12 electrons.  The sets I and II make
approximately $1.9\times 10^{-5}$ and $3.8\times 10^{-5}$ of the
complete basis.  The diamonds show the difference in correlation
function $G(\bbox{\rho})$ for two equivalent WC vectors (3,0) and
(0,6).

FIG.\ 4.  Sensitivity $S(\Phi)$ as a function of $\Phi$ at $J=0.04$
for two different systems.  Note the strong admixture of the second
harmonic, especially in the system $6\times 12$.


\begin{references}
\bibitem{dag}E. Dagotto, Rev. Mod. Phys. {\bf 66}, 763 (1994).
\bibitem{1d} J. des Cloizeaux, J. Math. Phys. {\bf 7}, 2136 (1966),
  B. S. Shastry, B. Sutherland, \prl {\bf 65}, 243 (1990).
\bibitem{pik} F. G. Pikus, A. L. Efros, Solid State Commun. {\bf 92},
  485 (1994), ibid. {\bf 96}, 183 (1995).
\bibitem{krive} Krive I. V. {\it et al.} Physica Scripta T, {\bf 54},
  123 (1994). 
\bibitem{kagan} Yu. Kagan, L. A. Maksimov, Sov. Phys. JETP
{\bf 38}, 307 (1974).
\bibitem{els} Our data for system $6\times 6$ with 7 electrons also
  show localization at small $J$.  These data will be published
  elsewhere.
\end{references}
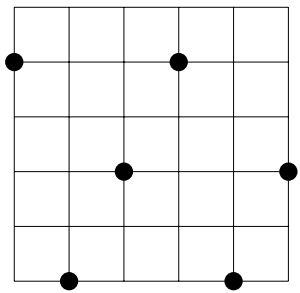
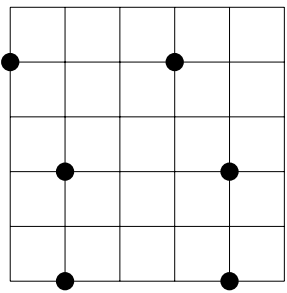
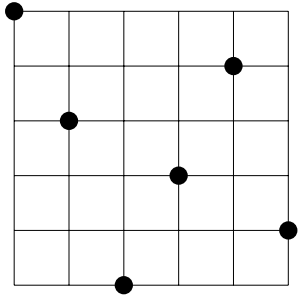
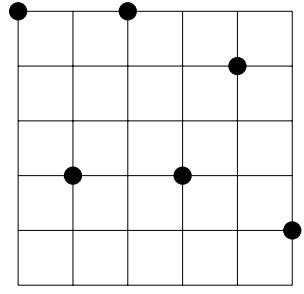
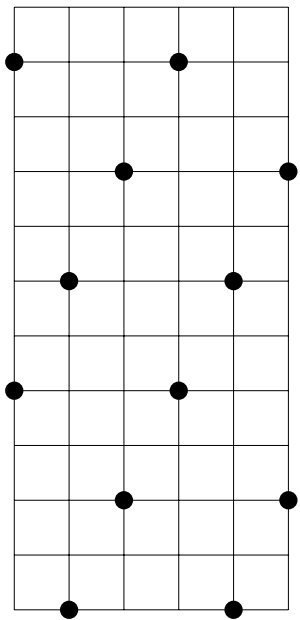
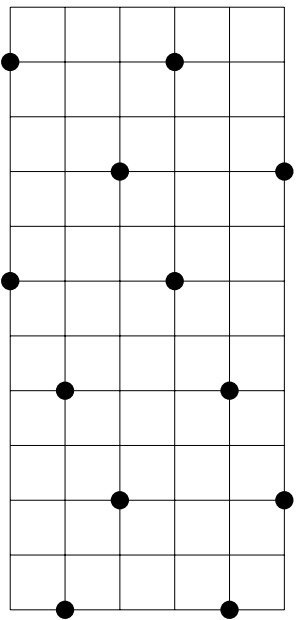
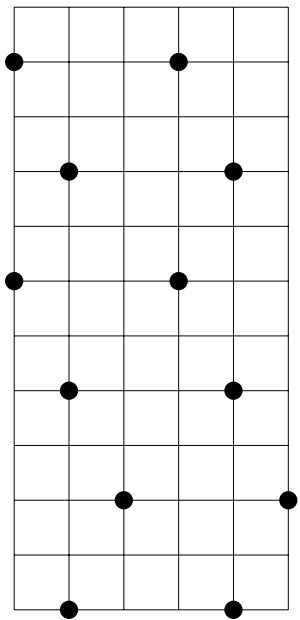
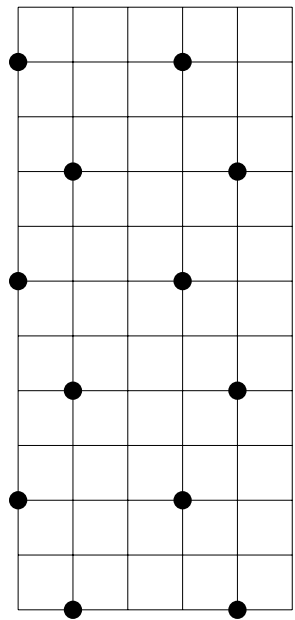
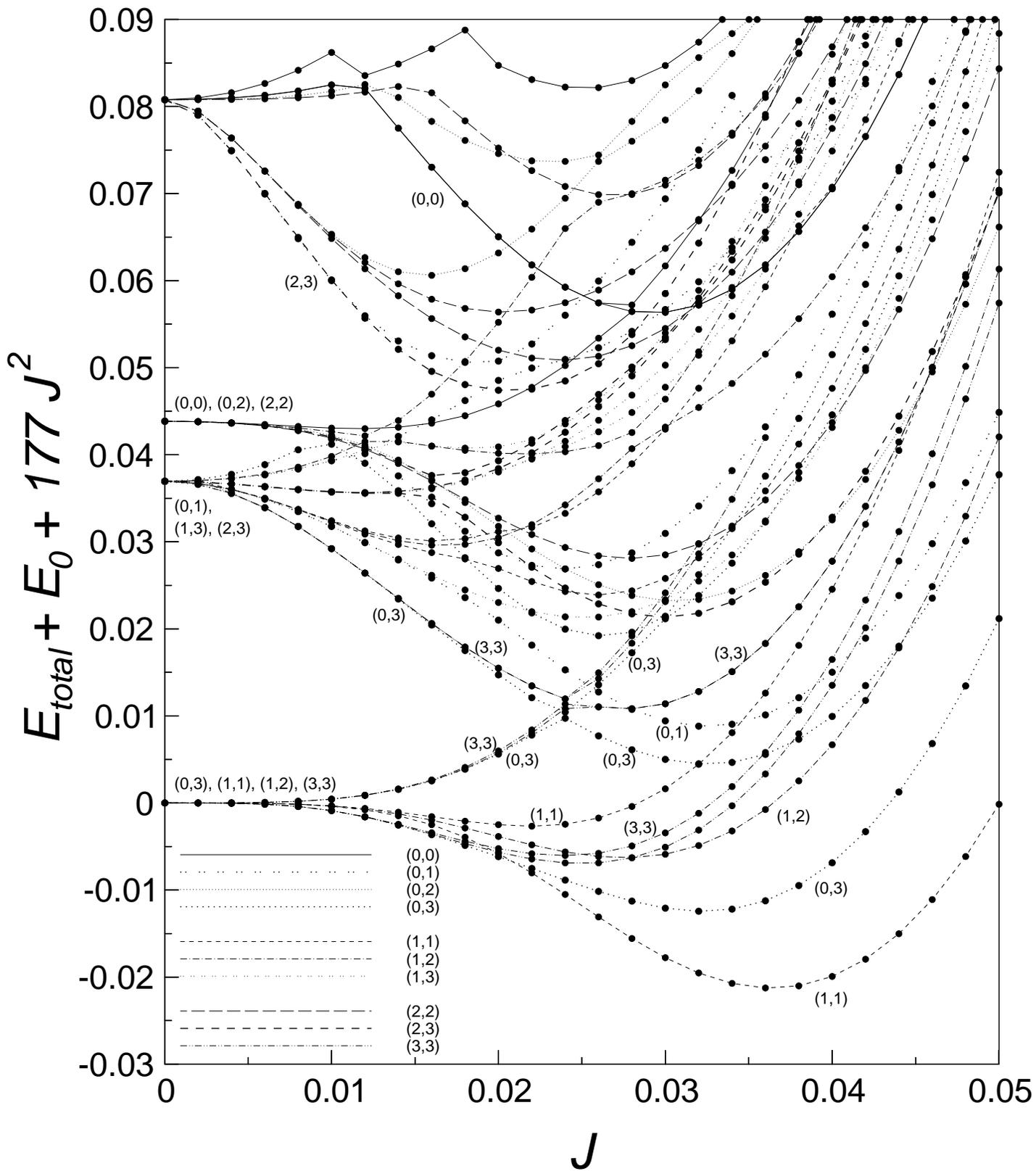
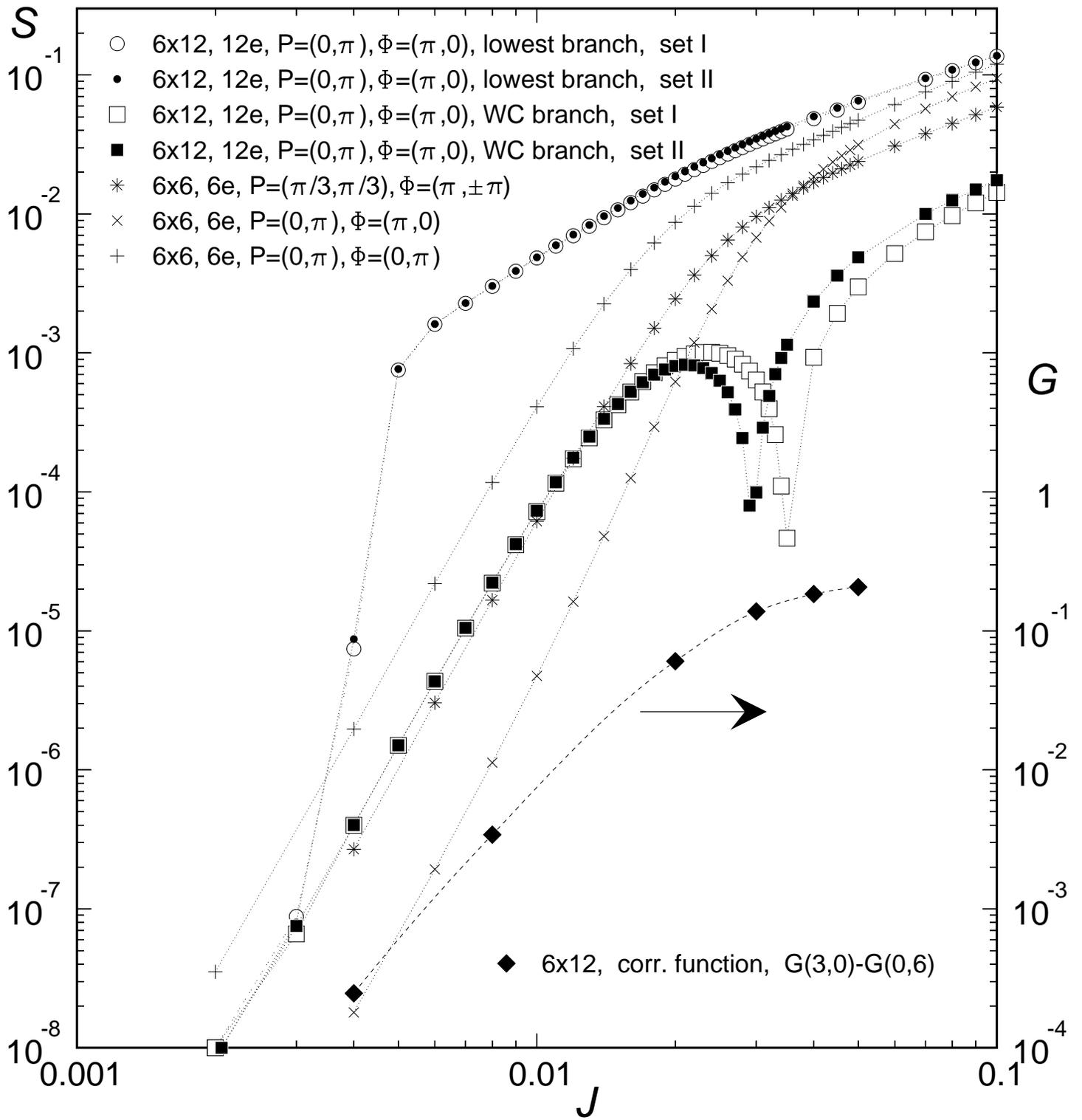
\end{document}